\documentclass[12pt]{article}
\usepackage[english]{babel}
\usepackage[cp1251]{inputenc}
\usepackage{geometry}
\usepackage{mathtext}
\usepackage[dvips]{graphicx}
\newcommand{\be}{\begin{equation}}
\newcommand{\ee}{\end{equation}}
\renewcommand{\baselinestretch}{1.4}
\date{}
\begin{document}
\title{High intensity generation of entangled photons in a two-mode electromagnetic field}
\normalsize
\author{D. N. Makarov \\
Northern (Arctic) Federal University,\\
163002, Arkhangelsk, Russia\\}
\maketitle
\begin{abstract}
At present, the sources of entangled photons have a low rate of photon generation. This limitation is a key component of quantum informatics for the realization of such functions as linear quantum computation and quantum teleportation. In this paper, we propose a method for high intensity generation of entangled photons in a two-mode electromagnetic field. On the basis of exact solutions of the Schr{\"o}dinger equation, in the conditions of interaction of the electrons in an atom with a strong two-mode electromagnetic field, it is shown that there may be large quantum entanglement between photons. The quantum entanglement is analyzed on the basis of the Schmidt parameter. It is shown that the Schmidt parameter can reach very high values depending on the choice of characteristics of the two-mode fields. We find the Wigner function for the considered case. Violation of Bell's inequalities for continuous variables is demonstrated.
\end{abstract}

\begin{center}
{\bf 1. Introduction}
\end{center}
Currently, the main method of producing quantum entangled states of light is spontaneous parametric light scattering (SPSL) \cite{Burnham_1970, Kwiat_1995, Fulconis_2007}. In this process, in the irradiation of a nonlinear doubly refractive crystal by the pumping field, with a certain low probability, the pumping photons split into two quantum-entangled photons of lower frequencies. In addition, the entangled pairs of photons can be obtained experimentally in the cascade decays in atomic systems  \cite{Aspect_1981}and in the parametric processes in the resonant fluorescence where, in the scattering, two pumping photons give birth to two quantum-entangled photons. Theoretical predictions of such photons \cite{Apanasevich_1979} have been confirmed in the experiment  \cite{Aspect_1980}. It should be noted that the entangled states can be not only for photons, but also for massive particles  \cite{Hagley_1997}.  The search for new sources of entangled photons is a topical problem. For example, in  \cite{Young_2006, Muller_2009} the quantum entanglement in radiative decay of two electron-hole pairs in the trap at a semiconductor quantum dot is studied. It should be noted that the sources of entangled photons have a very low generation rate of such photons. This is primarily due to the very low probability of formation of entangled states of photons. This limitation is a key component of quantum informatics for the realization of the functions such as linear quantum computations and quantum teleportation  \cite{Dousse_2010, Korzh_2015}. In this regard, the emergence of new ideas and methods with a high level of generation of entangled states is a topical problem.

In this paper, we propose a method of high intensity generation of entangled states of photons by a two-mode electromagnetic field. The source of these fields can be, for example, two intersecting single-mode laser beams which interact with electrons in atoms. As will be shown, in the course of this interaction, there may appear strongly entangled states of photons. An important part of the work is an exact analytical solution of the Schr{\"o}dinger equation for the interaction of an electron in an atom with a strong quantized two-mode electromagnetic field. In the semi-classical theory, this solution corresponds to the exact solution of the self-consistent system of Maxwell equations for the classical electromagnetic field and the Schr{\"o}dinger equation for an electron in an atom. Using the obtained solution, one can explicitly identify the parameters, for which strong quantum entanglement takes place, and also determine the characteristics of the two mode field. To quantify the degree (measure) of entanglement of photons, we will use the Schmidt parameter \cite{Grobe_1994, Ekert_1995}. It should be said that the Schmidt parameter is often used in the determination of the measure of quantum entanglement and the analysis of entangled states. It is known that a shortcoming of any measure of entanglement is the impossibility of direct experimental detection. Therefore, we introduce another measure of entanglement having a physical meaning, which in the end will coincide with the Schmidt parameter. We will also find the Wigner function, which can be used directly in the experiment, as well as in checking the violation of Bell's inequalities for continuous variables \cite{Samuel_1998, Jeong_2003, Bell_1964}.

\begin{center}
{\bf 2. Formulation of the problem. Solution of the Schr{\"o}dinger equation}
\end{center}

Consider the interaction of a two-mode electromagnetic field with an electron in an atom; in this case, the Schr{\"o}dinger equation will have the form
\begin{equation}
i\hbar \frac{\partial \Psi}{\partial t}=\left\lbrace \frac{1}{2m_{e}}\left( -i\hbar \frac{\partial }{\partial {\bf {r}}}+\frac{e}{c}{\hat{\bf{ A}}}\right)^2 + {\hat H}_{f} + U({\bf {r}}) \right\rbrace \Psi.
\label{1}
\end{equation}
In the expression (\ref{1}) $m_{e}$, -$e$ are, respectively the mass and charge of the electron, $U({\bf {r}})$ is the atomic potential, ${\hat H}_{f}=\sum_{\bf {k},\bf {u}}\hbar \omega \left( {\hat a}^{+}_{\bf {k},\bf {u}} {\hat a}_{\bf {k},\bf {u}}+\frac{1}{2}\right) $  is the Hamiltonian of the electromagnetic field, where  ${\hat a}^{+}_{\bf {k},\bf {u}}$ and ${\hat a}_{\bf {k},\bf {u}}$ are the operators of creation and annihilation of photons with the wave vector ${\bf k}$ and polarization  $\bf {u}$, whereas the vector potential  ${\hat{\bf{ A}}}$ will be
\begin{equation}
{\hat{\bf{ A}}}=\sum_{\bf {k},\bf {u}} \sqrt{\frac{2\pi c^2 \hbar }{\omega V_{f}}}\left({\bf {u}}exp\left(i(\omega t - {\bf k r}) \right){\hat a}_{\bf {k},\bf {u}} + {\bf {u}}^*exp\left(-i(\omega t - {\bf k r}) \right){\hat a}^{+}_{\bf {k},\bf {u}} \right) .
\label{2}
\end{equation}
Summation for ${\hat H}_{f}$ and in (\ref{2}) is performed over all possible values of the wave vector ${\bf k}$ and polarization $\bf {u}$. Next, we consider a field that will be called a two-mode electromagnetic field; in such a field,  $\sum_{\bf {k},\bf {u}} $ should be replaced by the sum of two terms, where for the first mode of the field we will have the wave vector ${\bf k}_{1}$ with polarization ${\bf u}_{1}$, whereas for the second mode, the wave vector ${\bf k}_{2}$ with polarization ${\bf u}_{2}$.  Next, for convenience, we proceed to use the atomic system of units  $\hbar=1, m_{e}=1, e =1$. Now consider the expression  (\ref{1})in the dipole approximation. In this approximation, the vector potential for a two-mode electromagnetic field expressed in terms of the field variables will be 
\begin{equation}
{\hat{\bf{ A}}}=a_{1}{\bf u}_{1} q _{1}+a_{2}{\bf u}_{2} q _{2},
\label{3}
\end{equation}
where $a_{1}=\sqrt{\frac{4\pi c^2}{\omega_{1}V_{1}}}$, $a_{2}=\sqrt{\frac{4\pi c^2 }{\omega_{2}V_{2}}}$, while  $\omega_{1},\omega_{2}$ are the frequencies of the first and second modes, respectively (we will conventionally assume that  $\omega_{2}>\omega_{1}$ if the frequencies of fields are different), $V_{1}, V_{2}$ are the quantization volume of the first and second modes, respectively, whereas  $q _{1}, q _{2} $ are the field variables of the first and second modes, respectively. As a result, the expression (\ref{1}) becomes
\begin{eqnarray}
i\frac{\partial \Psi}{\partial t}=\Biggl\lbrace \frac{1}{2}\left( -i \frac{\partial }{\partial {\bf {r}}}+\beta_{1}{\bf u}_{1} q _{1}+\beta_{2}{\bf u}_{2} q _{2}\right)^2+\frac{\omega_{1}}{2}\left(q^2_{1}-\frac{\partial^2 }{\partial {q_{1}^2}} \right)+\frac{\omega_{2}}{2}\left(q^2_{2}-\frac{\partial^2 }{\partial {q_{2}^2}}\right)  + U({\bf {r}}) \Biggr\rbrace \Psi ,
\label{4}
\end{eqnarray}
where $\beta_{1}=\sqrt{\frac{4\pi }{\omega_{1}V_{1}}}$, $\beta_{2}=\sqrt{\frac{4\pi}{\omega_{2}V_{2}}}$. Further, we will assume that the electromagnetic field is so strong that the potential $U({\bf {r}})$can be neglected. As a result, we need to solve the following stationary Schr{\"o}dinger equation ${\hat{H}}\Psi (q_{1},q_{2},{\bf r})=E \Psi (q_{1},q_{2},{\bf r})$, where 
\begin{eqnarray}
{\hat{H}}= \frac{1}{2}\left( -i \frac{\partial }{\partial {\bf {r}}}+\beta_{1}{\bf u}_{1} q _{1}+\beta_{2}{\bf u}_{2} q _{2}\right)^2 + \frac{\omega_{1}}{2}\left(q^2_{1}-\frac{\partial^2 }{\partial {q_{1}^2}} \right)+\frac{\omega_{2}}{2}\left(q^2_{2}-\frac{\partial^2 }{\partial {q_{2}^2}}\right)  ,
\label{5}
\end{eqnarray}
whereas the function  $\Psi (q_{1},q_{2},{\bf r},t)=e^{-i{\hat{H}}t}\Psi (q_{1},q_{2},{\bf r})$. The solution of the stationary Schr{\"o}dinger equation with Hamiltonian (\ref{5}) has the form (the decision in Appendix A) 
\begin{eqnarray}
\Psi_{{\bf k},m,n}= C_{\bf k} C_{n} C_{m}e^{i \lambda_{2}k_{z}x + i{\bf k}{\bf r}}exp\left( -\frac{\Lambda}{2L}\left(q_{2}+\lambda_{1}k_{z}+\xi_{2}k_{x}+\eta q_{1} \right) ^2\right) \times
\nonumber\\ 
H_{n}\left(\left(q_{2}+\lambda_{1}k_{z}+\xi_{2}k_{x}+\eta q_{1} \right)\sqrt{\frac{\Lambda}{L}} \right)exp\left( -\frac{W}{2V}\left( q_{1}(1-\xi \eta)+\xi_{1}k_{z}-\xi q_{2}\right)^2\right)
\nonumber\\ 
\times H_{m}\left(\left( q_{1}(1-\xi \eta)+\xi_{1}k_{z}-\xi q_{2}\right)\sqrt{\frac{W}{V}} \right). 
\label{20}
\end{eqnarray}

It can be seen from the expression (\ref{20}) that the states of the system described by the Hamiltonian (\ref{5}) cannot be expressed in terms of the states with a given number of "pure" photons. In addition, the obtained result cannot be expressed through a state of "dressed" photons of the single-mode electromagnetic field \cite{Gonoskov}. Also, the general form of the expression (\ref{20}) is such that it is impossible to single out any individual particle, and therefore all the particles of the system interact with each other. It is exactly this circumstance that leads to quantum entanglement of the system under consideration. It is necessary to clarify the kind of quantum entanglement in question. If we consider the case when the parameters $\xi$ and $\eta$ equal zero (for example, the polarizations ${\bf u}_{1}{\bf u}_{2} =0$ are orthogonal), then it is seen from (\ref{20}) that quantum entanglement between the photons disappears, but there is entanglement between the photons of the considered two-mode fields and the electrons in the atom. If we consider the case when the bond between the electron and the nucleus in the atom is too weak (the Rydberg atom), then in the expression (\ref{20}) $k_{x}$ and $k_{z}$ are small quantities, which leads to quantum entanglement only between the photons of the two-mode field (elastic scattering). The general case leads to quantum entanglement between the system as a whole: the electron and photons of the two-mode field. In the present work, we analyze only the quantum entanglement between the photons of the two-mode field.

Using the wave function (\ref{20}) and Hamiltonian (\ref{15}), it is not difficult to obtain an expression for  $\Psi (q_{1},q_{2},{\bf r},t)$ in terms of the sum 
\begin{eqnarray}
\Psi (q_{1},q_{2},{\bf r},t)=\sum_{n,m,{\bf k}}A^{s_1,s_2,0}_{m,n,{\bf k}}\Psi_{{\bf k},m,n}e^{-i E_{{\bf k},m,n}t} ,
\label{21}
\end{eqnarray}
$A^{s_1,s_2,0}_{m,n,{\bf k}}$ is the amplitude of transition from the initial states  $|s_1, s_2, 0 \rangle = |s_1\rangle |s_2\rangle |0\rangle$ of the system to the final one  $| \Psi_{{\bf k},m,n} \rangle $, where $|s_1\rangle,|s_2\rangle$ are the initial states of the field with one and two modes, respectively, and  $|0\rangle$ is the initial state of the electron in the atom. Amplitude $A^{s_1,s_2,0}_{m,n,{\bf k}}$ has the form 
\begin{eqnarray}
A^{s_1,s_2,0}_{m,n,{\bf k}}=\langle \Psi_{{\bf k},m,n} |  s_1, s_2, 0  \rangle  .
\label{22}
\end{eqnarray}
An analogue of the obtained solution  (\ref{21}) in the semi-classical theory  \cite{Scully_1997} is the exact solution of the self-consistent system of Maxwell's equations and the Schr{\"o}dinger equations for a charge in a strong electromagnetic field. Obviously, in the semi-classical physics, this problem is difficult, even with the use of numerical methods of calculation. The uniqueness of the solution  (\ref{21}) is connected with the fact that such a solution is found in the quantum physics, and it can be used not only for the analysis of quantum entanglement.

Next, we make a simplification, connected with the fact that the parameters $\beta_{1}, \beta_{2}$ in the expressions (\ref{17}) and (\ref{20}) are small quantities. Indeed, for a realistic microcavity or focal volume \cite{Tey_2008}, this quantity assumes the values of the order of $10^{-5} - 10^{-3}$,but it is usually much smaller than even these values. It can be seen from the expressions (\ref{13}) and (\ref{20}) that the entanglement of photons will be significant in the case when the parameter $\epsilon=\alpha/\gamma$ is a finite quantity. Since  $\gamma \sim \beta_{1} \beta_{2} <<1$, while  $\epsilon$ is a finite quantity, we get that  $\alpha<<1$.   As a result, it can be seen from (\ref{12}) that $\alpha<<1$ only when  $\omega_{1} \approx \omega_{2}\approx  \omega$,  while  $\Delta \omega = \omega_{2}-\omega_{1}$ is less or of the order of  $\beta_{1}\beta_{2}$.
Let us write the expressions (\ref{17}) and (\ref{20}) while retaining their main terms under the condition $\gamma \sim \beta_{1} \beta_{2} <<1$ and $\omega_{1} \approx \omega_{2}\approx  \omega$
\begin{eqnarray}
E_{{\bf k},m,n}=\frac{k^2}{2}+\omega_{1}( m+\frac{1}{2}) +\omega_{2}( n+\frac{1}{2})+\frac{m}{2}\left(\beta^2_{1}-\eta \beta_{1}\beta_{2}{\bf{u}}_{1} {\bf{u}}_{2} \right)+\frac{n}{2}\left(\beta^2_{2}+\eta \beta_{1}\beta_{2}{\bf{u}}_{1} {\bf{u}}_{2} \right),  ~~~
\label{23}
\end{eqnarray}
\begin{eqnarray}
\Psi_{{\bf k},m,n}= C_{\bf k} C_{n} C_{m}e^{i{\bf k}{\bf r}}exp\left( -\frac{\sigma}{2}\left(q_{2}+\eta q_{1}+\lambda_{1}k_{z}+\xi_{2}k_{x} \right) ^2\right) \times
\nonumber\\ 
H_{n}\left(\sqrt{\sigma}\left(q_{2}+\eta q_{1} +\lambda_{1}k_{z}+\xi_{2}k_{x}\right) \right)exp\left( -\frac{\sigma}{2}\left( q_{1}-\eta q_{2}+\frac{\xi_{1}k_{z}}{\sigma}\right)^2\right)
\nonumber\\ 
\times H_{m}\left(\sqrt{\sigma}\left( q_{1}-\eta q_{2}+\frac{\xi_{1}k_{z}}{\sigma}\right) \right),
\label{24}
\end{eqnarray}
where $\sigma = \frac{1}{1+\eta^2}$ while, according to (\ref{13}) $\eta=\frac{{\bf u}_1{\bf u}_2}{|{\bf u}_1{\bf u}_2|}(\sqrt{1+\epsilon^2}-|\epsilon|)$. The parameter $\eta $ is a smooth function, since for ${\bf u}_1{\bf u}_2=0$ the parameter  $\eta=0$ and its values can be in the range $\eta\in (-1,1) $,  thus   $\sigma \in (1/2,1)$ . It should be noted that in the expression (\ref{24}) $\lambda_{1}, \xi_{1}, \xi_{2}$ are small quantities  $\sim (\beta_{1}, \beta_{2})$. In spite of this, $\lambda_{1}, \xi_{1}, \xi_{2}$ are responsible for the inelastic processes, which can take place during the interaction of the two-mode field with the electron in the atom. Therefore, these terms, in spite of their smallness, are essential. Indeed, if we disregard the elastic processes and consider a single-mode field, then they can make a significant contribution to the photoionization \cite{Gonoskov}. It should also be said that these terms increase or decrease the number of photons in the system, because they are responsible for inelastic processes (this will be explicitly shown below). Using the wave function (\ref{24}), we can obtain the amplitude  (\ref{22}) in an analytical form (the details of calculation are in the Appendix B)
\begin{eqnarray}
A^{s_1,s_2,0}_{m,n,{\bf k}}=F_{{\bf k},0}\sum^{s_{1}+s_{2}}_{p=0}G^{In}_{m,n,s_1+s_2 - p,p}({\bf k})G^{El}_{s_1+s_2-p,p,s_{1},s_{2}} ,
\label{25}
\end{eqnarray}
where 
\begin{eqnarray}
G^{El}_{k,p,s_{1},s_{2}}=\frac{\eta^{s_{1}+k}\sqrt{p!k!}}{(1+\eta^2)^{\frac{s_1+s_2}{2}}\sqrt{s_{1}!s_{2}!}}P^{(-(1+s_1+s_2), p-s_1)}_{k}\left(-\frac{2+\eta^2}{\eta^2} \right) ,
\label{26}
\end{eqnarray}
\begin{eqnarray}
G^{In}_{m,n,k,p}({\bf k})=\frac{(-1)^{(p-n)\theta(p-n)}(-1)^{(k-m)\theta(k-m)}}{\sqrt{m!n!p!k!}}e^{-\frac{1}{4}\left(\frac{(\xi_{1}k_{z})^2}{\sigma}+  (\lambda_{1}k_{z}+\xi_{2}k_{x} )^2\sigma \right) } 
\nonumber\\ 
\times Dn!Dm! \left(\frac{\xi_{1}k_{z}}{\sqrt{2\sigma}} \right)^{|m-k|} \left( (\lambda_{1}k_{z}+\xi_{2}k_{x} )\sqrt{\frac{\sigma}{2}}\right)^{|n-p|} 
\nonumber\\ 
\times L^{|n-p|}_{Dn}\left( (\lambda_{1}k_{z}+\xi_{2}k_{x} )^2\frac{\sigma}{2}\right)L^{|m-k|}_{Dm}\left(\frac{(\xi_{1}k_{z})^2}{2\sigma} \right)
,
\label{27}
\end{eqnarray}
\begin{eqnarray}
F_{{\bf k},0}=C_{\bf k}\langle e^{i{\bf k}{\bf r}}|0\rangle .
\label{28}
\end{eqnarray}
In the expression (\ref{26}) $P^{(b,c)}_{a}(x)$ are the Jacobi polynomials, in (\ref{27}) $\theta(x)$ is the Heaviside theta-function,  $L^{b}_{a}(x)$ is the generalized Laguerre polynomial, $Dm = (m+k-|m-k|)/2$, $Dn = (n+p-|n-p|)/2$. As shown in the Appendix B, we should take into account in the expressions (\ref{26}) and (\ref{27}) that $k+p=s_1+s_2$ (the number of particles in elastic scattering is preserved).

The expression (\ref{25}) can be analyzed rather simply, despite of the fact that $G^{El}$ and $G^{In}$ are complicated function. Indeed, if we consider only elastic processes, the function  $G^{In}=1$ (the number of particles in the system remains the same $m+n=s_1+s_2$), while in the sum of expression (\ref{25}) there remains one term with  $p=n$. If scattering processes are unlikely and can be neglected, then $G^{El}=1$, in this case, one term with $p=s_2$  remains in the sum of expression (\ref{26}). Thus, in the expression (\ref{25}) $G^{El}$ is responsible for the elastic processes, whereas  $G^{In}$, for the inelastic processes. As a result, on the basis of the expression (\ref{25}), we can evaluate the processes in the considered problem. The analysis of the processes can be carried out efficiently, if we see that in the expression  (\ref{25}) these processes proceed successively. If we consider such process when in the first and second modes of the field there were $s_1$ and $s_2$ photons, respectively, it turns out that some of them are scattered, for example, $k$ photons from the mode $s_2$, then in the mode $s_2$ after scattering we have  $p=s_2-k$ photons, whereas in the mode $s_1$, there is an increase by $k$ photons, thus in this mode there is now $s_1+k=s_1+s_2-p$ this process is described exactly by the matrix element  $G^{El}_{s_1+s_2-p,p,s_{1},s_{2}}$. It should be noted that the number of particles in the scattering, described by the matrix element $G^{El}_{s_1+s_2-p,p,s_{1},s_{2}}$ is always preserved and equals $s_1+s_2$. Further, in the fields in which there remains for the first mode $s_1+s_2-p$ photons, and for the second, $p$ photons, there take place inelastic processes in which the number of photons in each mode may increase or decrease. As a result, there may remain in the first mode  $m$ and in the second mode $n$ photons; and this process is described by the matrix element $G^{In}_{m,n,s_1+s_2 - p,p}$. It should be noted that, for the values of $|\eta|$ close to one, such an analysis is rather provisory. Indeed, the scattering depends on the parameter $\eta$, but $G^{In}$, also depends on this parameter, although for  $|\eta|$ close to one this dependence is not strong enough to qualitatively change the picture. This is due to the fact that, for  $|\eta|$ close to one, the entire system is the most entangled. If we use the terminology of semi-classical scattering theory, it is possible to interpret this process as self-consistent radiation.

\begin{center}
{\bf 3. Quantum entanglement }
\end{center}

According to the Schmidt theorem, the wave function  $\Psi (q_{1},q_{2},{\bf r},t)$ can be decomposed with respect to the Schmidt modes in the form
\begin{eqnarray}
\Psi (q_{1},q_{2},{\bf r},t)=\sum_{m_{1},m_{2},m_{3}} \sqrt{\Lambda_{m_{1},m_{2},m_{3}}(t)}\Phi_{m_{1}}(q_{1})\Phi_{m_{2}}(q_{2})\phi_{m_{3}}({\bf r}), 
\label{29}
\end{eqnarray}
where $\Phi_{m_{1}},\Phi_{m_{2}}$ are the Schmidt modes of the non-interacting electromagnetic field, $\phi_{m_{3}}({\bf r})$ is the Schmidt mode for the non-interacting electron in the atom. It is known that  $\sum \Lambda_{m_{1},m_{2},m_{3}}(t)=1$ and $0<\Lambda_{m_{1},m_{2},m_{3}}(t)<1$. A quantitative measure of entanglement is the Schmidt parameter 
\begin{eqnarray}
K(t)=\frac{1}{\sum_{m_{1},m_{2},m_{3}} \Lambda^2_{m_{1},m_{2},m_{3}}(t)}.
\label{30}
\end{eqnarray}
It is known that the Schmidt parameter $K$ is a quantitative characteristic of the number of such terms in (\ref{29}) that are not small in this sum. Of course, this measure of entanglement is purely theoretical, suitable for analysis of entangled states. In this case, the Schmidt parameter is a quite complicated parameter for calculation, because the considered problem is a many-particles one. Analogously to the Schmidt parameter, further in the work we will introduce a measure of entanglement which is more understandable from the physical point of view, simpler for calculation and numerically very close to the Schmidt parameter. Using (\ref{29}), we obtain
\begin{eqnarray}
\Lambda^{{s_{1},s_{2},0}}_{m_{1},m_{2},m_{3}}(t)=\Biggr|\sum_{n,m,{\bf k}}A^{s_1,s_2,0}_{m,n,{\bf k}}A^{* m_{1},m_{2},m_{3}}_{m,n,{\bf k}}e^{-i E_{{\bf k},m,n}t}\Biggl|^2.
\label{31}
\end{eqnarray}
We are interested in the quantum entanglement of photons, so we sum the parameter $\Lambda^{{s_{1},s_{2},0}}_{m_{1},m_{2},m_{3}}(t)$, over all electron states  $m_{3}$ and, finally, we obtain the desired parameter in the form 
\begin{eqnarray}
\Lambda^{{s_{1},s_{2},0}}_{m_{1},m_{2}}(t)=\sum_{\bf k}|F_{{\bf k},0}|^2\Biggr|\sum_{n,m}\sum^{s_1+s_2}_{(p_{1},p_{2})=0}G^{In}_{m,n,s_1+s_2 - p_1,p_1}({\bf k})G^{El}_{s_1+s_2-p_1,p_1,s_{1},s_{2}}
\nonumber\\ 
\times G^{In}_{m,n,m_1+m_2 - p_2,p_2}({\bf k})G^{El}_{m_1+m_2-p_2,p_2,m_{1},m_{2}}e^{-i E_{m,n}t}\Biggl|^2 ,
\label{32}
\end{eqnarray}
where $ E_{m,n}$ is defined by expression (\ref{23}), but without the kinetic energy $k^2/2$ of the electron. It should be noted that, using expression (\ref{32}), we obtain the Schmidt parameter for the entanglement of photons between themselves and between the electron. Consider first the elastic processes  $G^{In}=1$, where the quantum entanglement will be between the photons only. In this case, the parameter $\Lambda^{{s_{1},s_{2},0}}_{m_{1},m_{2}}(t)$ will be significantly simplified and equal to 
\begin{eqnarray}
\Lambda^{{s_{1},s_{2},0}}_{m_{1},m_{2}}(t)=\Biggr|\sum^{s_1+s_2}_{n=0}G^{El}_{s_1+s_2-n,n,s_{1},s_{2}}G^{El}_{m_1+m_2-n,n,m_{1},m_{2}}e^{-i n \delta t}\Biggl|^2 ,
\label{33}
\end{eqnarray}
where $ \delta=\Delta \omega +1/2(\beta^2_{2}-\beta^2_{1})+\eta \beta_{1} \beta_{2}{\bf{u}}_{1} {\bf{u}}_{2} $. When calculating the Schmidt parameter, we should take into account that $m_1+m_2=s_1+s_2$ (the number of particles is preserved). An interesting case is the one not connected with the quantum entanglement: scattering on the zero vacuum fluctuations $s_2=0$. In the expression (\ref{33}), in the case of the weak interaction $|\eta|<<1$, and taking into account the smallest order of the parameter  $\eta$, we find that the probability of scattering  $W=s_1\beta^2_{1}\beta^2_{2}({\bf{u}}_{1} {\bf{u}}_{2})^2\frac{sin^2(\Delta \omega t)}{\Delta \omega^2}$. Since the obtained expression is linear with respect $\beta^2_{1}\beta^2_{2}$ we can sum over the entire phase volume of the second field and average over its polarization. If we take into account that $I^2_{1}V=8\pi \omega_{1}s_1$, which corresponds to  $\beta^2_1=I^2_{1}/(2\omega^2_1 s_1)$, where $I_{1}$ is the intensity amplitude of the first electromagnetic field, we get the Thomson formula. It is known that the quantum calculation of the scattering cross section and the classic one ($s_1\to \infty$) yield one and the same result: the Thomson formula for scattering. Here it is exactly demonstrated that the result does not depend on the number of photons  $s_1$.

Let us present the calculation results for the Schmidt parameter  $K(t)$ in the case of elastic scattering. The calculation results are presented in Fig. 1 and Fig. 2. It should be noted that we present in the figures the dependence on $\delta t$ from 0 to $2\pi$, which is quite justified, because this dependence is repeated with the period of $2\pi$ (it can be seen from formula  (\ref{33}) ). To characterize the quantum entanglement, it suffices to know the average value of the Schmidt parameter. To this end, we average its value with respect to  $\delta t$ over the period. Consider the case of equal fields $\omega_1\approx \omega_2 \approx \omega, V_1\approx V_2=V, s_1=s_2=s $. Analyzing the average value of the Schmidt parameter for these fields, we can obtain the following formula for $K=\langle K(t)\rangle$
\begin{eqnarray}
K=\frac{1}{1+3\eta^{2.15}}\frac{ 114.5(\eta s)^{3}+1}{1+35(\eta s)^2},
\label{34}
\end{eqnarray}
where $\eta\equiv|\eta|$,because it can be seen from (\ref{33}) and (\ref{26}) that $(G^{El}(\eta))^2=(G^{El}(-\eta))^2$, thus, the result does not depend on the sign of $\eta$. In what follows, wherever we consider elastic scattering, we will assume $\eta\equiv|\eta|$.
\begin{figure}[!hbp]
\begin{center}
\includegraphics[angle=0,width=0.7\textwidth, keepaspectratio]{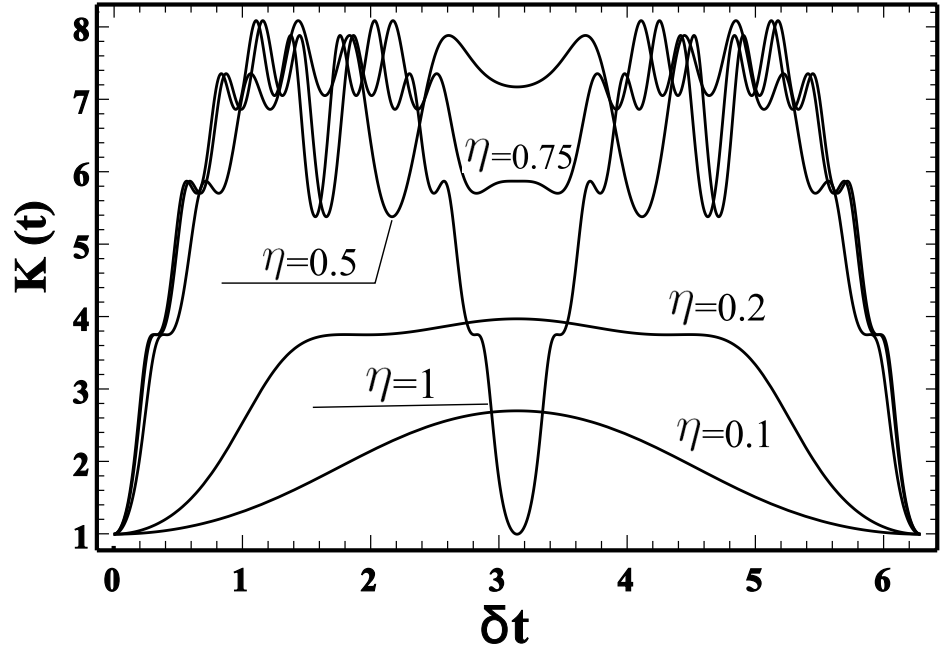}
\end{center}
\caption[fig_1]{The results of calculations of the Schmidt parameter $K(t)$ in its dependence on the dimensionless parameter  $\delta t$, for $s_1=s_2=5$ and $\eta=1,3/4,1/2,1/10,1/20$.}
\label{fig1}
\end{figure}
\begin{figure}[!hbp]
\begin{center}
\includegraphics[angle=0,width=0.7\textwidth, keepaspectratio]{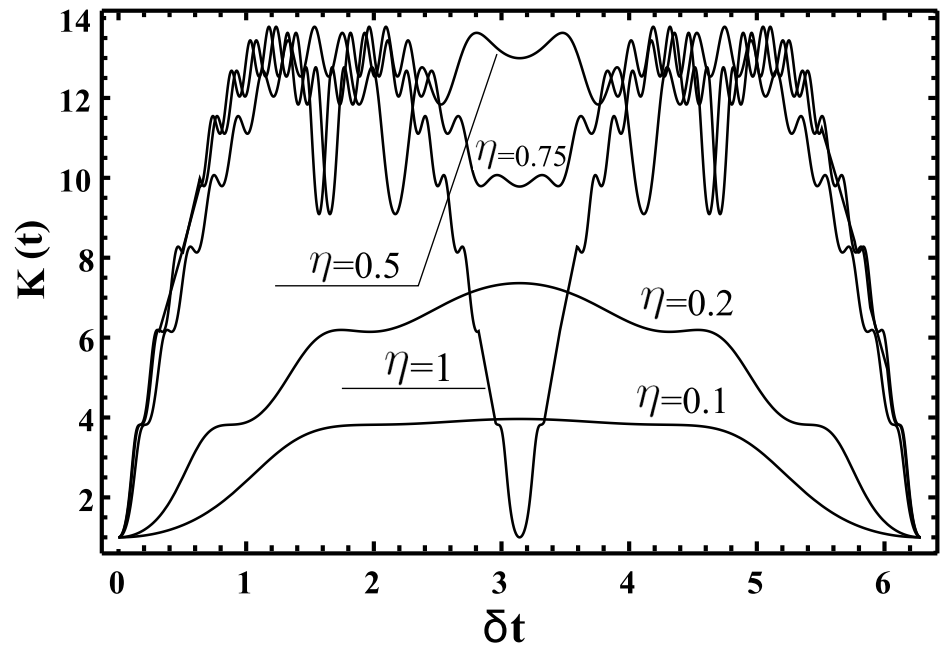}
\end{center}
\caption[fig_2]{The same as in Fig. 1, but for  $s_1=s_2=10$}
\label{fig1}
\end{figure}

It should be said that in the expression (\ref{34}) $s$ cannot be taken to be arbitrary large; this restriction is determined by that $\eta$ should be a finite quantity. The restriction with respect  $s$ is determined by the characteristics of the considered two-mode field. Indeed, $\eta=\sqrt{1+\epsilon^2}-|\epsilon|$, where  $\epsilon=\alpha/\gamma$. If we assume that $\Delta \omega=0$, then this dependence is mainly determined by the volume of the resonators 1 and 2 of the field, where $\epsilon=1/2(\sqrt{V_1/V_2}-\sqrt{V_2/V_1})$ ($min\{\epsilon\}=0$, hence,  $max\{\eta\}=1$); having chosen close sizes of the resonators, we see that $s$ is not limited by anything, even when $ V_1\approx V_2=V \to \infty$. Although in the reality we cannot choose  $\Delta \omega=0$ but we can select a very small value of it, while $s$ can be a very large quantity, and hence the quantum entanglement is also large. For example, for the characteristics of  Nd-YAG laser with the photon energy $\hbar \omega =1.17 eV$ and the intensity $10^{13}W/cm^2 $, the parameter $\eta$ will be close to one if we choose  $\Delta \omega/\omega\sim 10^{-8}, |{\bf{u}}_{1} {\bf{u}}_{2}|\sim 1, s\sim 10^8$, hence $K\sim 10^8$. The approximation (\ref{34}) is obtained by numerical analysis, the results of which indicate that for the final $\eta$, the parameter  $K$ approaches the asymptotic $K \sim s $. The numerical analysis was carried out for $s\in (1,1000)$, whereas the parameter  $\eta\in (1/1000,1)$. Let us present in Fig. 3 the dependence of the Schmidt parameter $K$ for various $\eta$. An interesting fact is that when the paramete $\eta$ is not small, then almost the entire quantum system is entangled because $K\sim s$.
\begin{figure}[!hbp]
\begin{center}
\includegraphics[angle=0,width=0.7\textwidth, keepaspectratio]{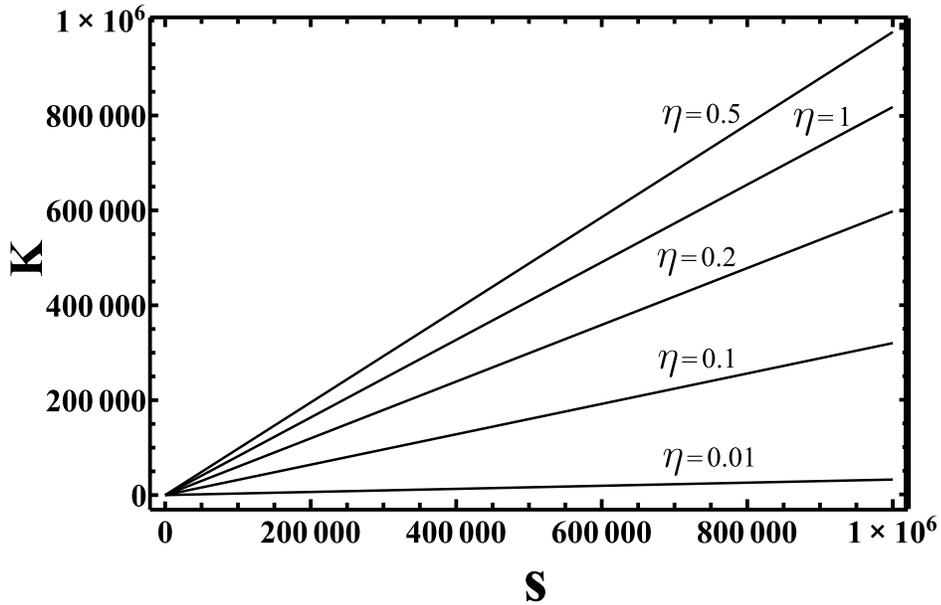}
\end{center}
\caption[fig_3]{The results of calculations of the Schmidt parameter $K$ in its dependence on the number $s$ of the quanta in the field for $\eta=1/100,1/10,1/5,1/2,1$.}
\label{fig3}
\end{figure}

Next, we consider the general case where the Schmidt parameter is defined by the expression (\ref{32}).Obviously, the inelastic scattering processes are described in a more complicated manner than the elastic ones. Therefore, we restrict ourselves to a qualitative analysis of the Schmidt parameter, and we show that quantum entanglement cannot be significantly reduced in comparison with the elastic processes. In the expression (\ref{27}), in the considered problem, the parameters $\frac{(\xi_{1}k_{z})^2}{2\sigma} $ and $(\lambda_{1}k_{z}+\xi_{2}k_{x} )^2\frac{\sigma}{2}$. are connected with changing the number of photons. It should be noted that these parameters are small, because $\lambda_{1},\xi_{1}, \xi_{2}<<1$. In the case of expression for  $G^{El}$, an analogous parameter responsible for the transitions (scattering) is the parameter $\eta$, which is a finite quantity. Hence, it can be concluded that the number of transitions for $G^{El}$ is larger than the number of quanta involved in the process $G^{In}$. If we consider the case where the numbers $s_1,s_2>>1$, it can be concluded that mostly the quantum entanglement is determined by $G^{El}$, hence, mainly it is the photons that are entangled with each other, rather than photons and the electron of the atom. Entanglement also depends on the considered atom; if it is the Rydberg one, then  $k_{z},k_{x}$, included in (\ref{27}), are small, and therefore  $G^{In}\to 1$.The issue of the precise calculation of quantum entanglement, taking into account of inelastic processes, is quite complicated and deserves a separate study, so further we limit ourselves to the quantum entanglement in the elastic scattering of photons. Next, we proceed to the consideration of the elastic scattering, i.e., $G^{In}\to 1$. 

For the analysis of quantum entanglement, it is convenient to introduce a parameter that has an obvious physical meaning: the average number  $N (t)$ of scattered particles. Define this parameter according to the definition of the average $N(t)=\sum^{s_1+s_2}_{m_1,m_2=0}\Lambda^{{s_{1},s_{2},0}}_{m_{1},m_{2}}(t)(|s_1-m_1|+|s_2-m_2|)$. Using the fact that the number of particles is conserved, i.e.,  $m_1+m_2=s_1+s_2$, we get
\begin{eqnarray}
N(t) =2\sum^{s_1+s_2}_{m_1=0}\Lambda^{{s_{1},s_{2},0}}_{m_{1},s_1+s_2-m_1}(t)|s_1-m_1|,
\label{36}
\end{eqnarray}
where $\Lambda^{{s_{1},s_{2},0}}_{m_{1},s_1+s_2-m_1}(t)$ is defined by the expression (\ref{33}) and is the probability of discovering the system at time $t$ in the state $|m_1\rangle | m_2\rangle$. Indeed, the expression  (\ref{36}) is a measure of quantum entanglement, because it determines the number of not small terms in the sum (\ref{29}),which is analogous to the Schmidt parameter. The expression  $N(t)$ besides the explicit physical meaning, has one more advantage over the Schmidt parameter: a more simple calculation. Indeed, it can be analytically averaged over time: $\langle N(t)\rangle =2\sum^{s_1+s_2}_{m_1=0}\sum^{s_1+s_2}_{n=0}\left( G_{s_1+s_2-n,n,s_1,s_2}^{El}G_{s_1+s_2-n,n,m_1,s_1+s_2-m_1}^{El}\right)^2|s_1-m_1| $. Consider the same case as for the Schmidt parameter: the fields of the first and second modes are equal, i.e.,  $\omega_1\approx \omega_2 \approx \omega, V_1\approx V_2=V, s_1=s_2=s $. Analyzing the numerical calculation of the average value and its time dependence, similarly to the Schmidt parameter, we obtain the following dependence for $\langle N(t)\rangle$. 
\begin{eqnarray}
\langle N(t)\rangle=\frac{1}{1+3\eta^{2.15}}\frac{ 114.5(\eta s)^{3}}{1+35(\eta s)^2},
\label{37}
\end{eqnarray}
The numerical analysis was carried out for $s\in (1,10000)$. It can be seen that the expressions (\ref{37}) and (\ref{34}) almost coincide; the difference between them is observed only for low  $\eta s$, if $\eta$   is close to one; so we can say that there is no difference. Indeed, for weak fields,  $K\to 1$, while  $\langle N(t)\rangle \to 0$. We are interested in the fields, where the entanglement is large, i.e, when  $\eta \sim 1$; for these values, the parameters tend to each other. 
We can qualitatively demonstrate the dependence of the Schmidt parameter $K(t)$ (hence $N(t)$)on time for the above case. This dependence is needed when the interaction time of the two-mode field with the electron in the atom is so small that it is impossible to average over $\delta t$. For the qualitative analysis, it is sufficient to assume that $K(t)\sim K \delta t$,  which is a linear approximation. In the considered case, when $\eta$ is not small, we have $K \sim \eta s$, while the parameter $ \delta=\Delta \omega +\eta \beta^2 {\bf{u}}_{1} {\bf{u}}_{2} $. We can assume that  $I^2V=8\pi \omega s $, which corresponds to  $\beta^2=I^2/(2\omega^2 s)$, where $I$ -is the intensity amplitude of the electromagnetic field. Then, for not small $\eta$, we obtain that $s \delta \sim I^2/\omega^2$, hence,  $K(t)\sim P \omega t$, where $P=I^2/(4\omega^3)$ is the quantum ponderomotive parameter, which is usually used in the photoionization theory \cite{Scully_1997, Reis_1992}. The parameter $P$ can be sufficiently large, whereas $\omega t$ can be chosen to be a large quantity, so  $K(t)>>1$ for small $\delta t$. Knowing $ N(t)$ possible to estimate the scattering cross section of the process. Since $N(t)\sim P \omega t \sim I^2 t/\omega^2$,the energy dissipation per unit time will $\frac{d \varepsilon}{dt} \sim I^2 /\omega$. The result is not difficult to obtain the scattering cross section $\sigma \sim 1/(c\omega)\sim \frac{c^3}{\omega}\sigma_{T}\sim \frac{10^6}{\omega}\sigma_{T}$, where $\sigma_{T}$ - Thomson scattering cross section. The results obtained in the qualitative analysis of the scattering cross section is very large value. The cross section generating entangled photons in this case is the huge value in comparison with known methods of generation of entangled photons. For an optical wavelength range, the scattering cross section will  $\sigma \sim 10^7 \sigma_{T}$. The considered cross-section of the scattering can be described as a cross section of quantum entanglement.

\begin{center}
{\bf 4. The Wigner function. Violation of Bell's inequality }
\end{center}
It is known that the function $W$ introduced by Wigner is a quasi-probability distribution and is used to determine various characteristics of the field \cite{Wigner_1932, Walther_1968, Mecklenbrauker_1997}. In addition, using this characteristic, it is possible to define Bell's inequality for continuous quantum variables \cite{Samuel_1998, Bell_1964, Banaszek_1999}. In our case,  $\Psi (q_{1},q_{2},{\bf r},t)$ in formula (\ref{21}) depends on 5 continuous variables  $q_{1},q_{2},{\bf r}$, but we are only interested in the field variables. Therefore, it is not difficult to conclude that the Wigner function in this case has the form
\begin{eqnarray}
W(x_1,x_2,p_1,p_2,t)= \frac{1}{(2\pi)^2}\int dq_1 dq_2 e^{i(q_1 p_1+q_2 p_2)}
\nonumber\\ 
\times 
\Psi^{*}(x_1+\frac{q_1}{2},x_2+\frac{q_2}{2},t)\Psi(x_1-\frac{q_1}{2},x_2-\frac{q_2}{2},t),
\label{41}
\end{eqnarray}
where $\Psi(q_1,q_2,t)$ is defined by the formula (\ref{21}) in which the electron wave function is equal to 1. Next, we consider only the elastic scattering, since for the inelastic processes the Wigner function becomes much more complicated and deserves special consideration in another work. Consider the Wigner function averaged over time, in the same way as it was done in the preceding paragraph. As a result, the Wigner function can be obtained in the analytical form
\begin{eqnarray}
W=\frac{(-1)^{s_1+s_2}}{\pi^2}\sum^{s_1+s_2}_{p=0}|G^{El}_{s_1+s_2-p,p,s_{1},s_{2}}|^2 L_{p}(X) L_{s_1+s_2-p}(Y)e^{-\frac{X}{2}}e^{-\frac{Y}{2}},
\label{42}
\end{eqnarray}
where $X=2\sigma \left((p_2+\eta p_1)^2+(x_2+\eta x_1)^2 \right) $, $Y=2\sigma \left((p_1-\eta p_2)^2+(x_1-\eta x_2)^2 \right) $, $L_{a}(z)$ is the Laguerre polynomial, $G^{El}_{s_1+s_2-p,p,s_{1},s_{2}}$ is the function calculated by the formula (\ref{26}). If we take $\eta=0$, in the expression  (\ref{42}), then the Wigner function will coincide with the known expression for the function $W$ in the Fock state, but taking into account that here we consider two fields. In the Fock state, the Wigner functions for the first and second fields are independent from each other, which is not true in the case $\eta \neq 0$. It is seen from equation  (\ref{42}) that, in general, $W$ is not symmetric with respect to the variables $x_1, x_2, p_1, p_2$. As an illustration, we present Fig. 4, where  $p_2=x_2=0$, then $W$ depends on one variable, the radius on the phase plane $r_1=\sqrt{x^2_1+p^2_1}$. It is seen from Figure 4 that the phase space is symmetrically compressed for not small $\eta $. The phase pattern will be even more complicated and not symmetric, if  $p_2,x_2\neq 0$. From Figure 4 it can be seen that for  $\eta =1/2$ the phase space is more compressed than for $\eta =1$. This can be easily explained if we analyze the maximum of expression (\ref{34}) and see that for $\eta =1$ the quantum entanglement is not maximal.
It is known \cite{Jeong_2003, Zhang_2007, Clauser_1969} that Bell’s inequalities for continuous variables have the following form
\begin{eqnarray}
B=\pi \left| W(a;b)+W(a;b^{'})+W(a^{'};b)-W(a^{'};b^{'})\right |<2 ,
\label{43}
\end{eqnarray}
where $a=(x_1,p_1), b= (x_2,p_2)$. Let us show that inequality (\ref{43}) may be violated. We select some values of $b^{'}$ and $p_1, p_2$, and demonstrate a violation of Bell’s inequalities  (\ref{43}) using the graph in Fig. 5. The Bell parameter $B$ may reach sufficiently large values, for example, for $p_1=p_2=p^{'}_1=p^{'}_2=0, x_1=-0.0971,x_2=0.126, x^{'}_1=0.00815,x^{'}_2=-0.0562,s_1=0,s_2=40$ the parameter $B=2.262$.

\begin{figure}[!hbp]
\begin{center}
\includegraphics[angle=0,width=0.7\textwidth, keepaspectratio]{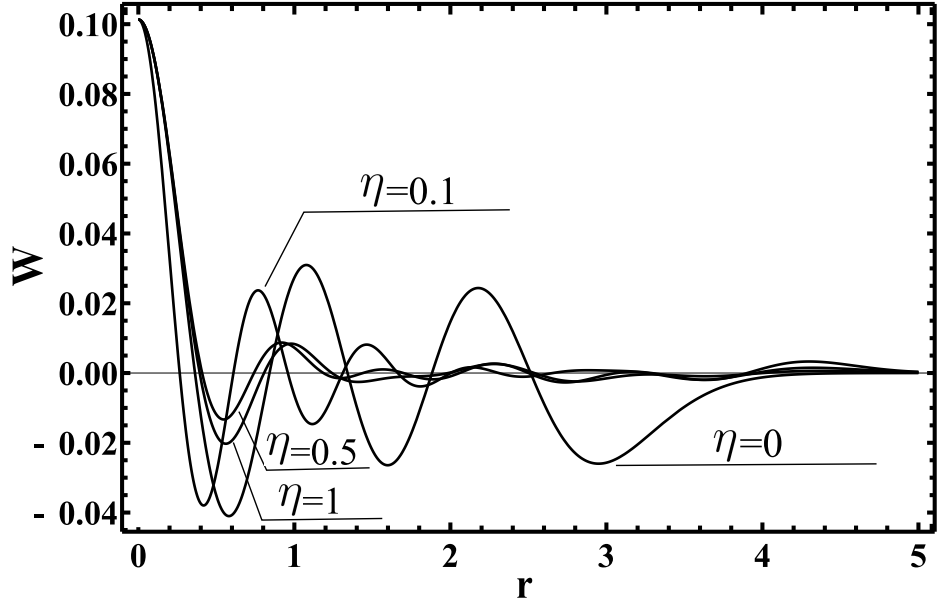}
\end{center}
\caption[fig_4]{The results of calculations of the Wigner function $W$ in its dependence on the radius  $r_1$ in the phase space for  $s_1=s_2=5$ and $\eta=0,1/10,1/2,1$}
\label{fig4}
\end{figure}
\begin{figure}[!hbp]
\begin{center}
\includegraphics[angle=0,width=0.7\textwidth, keepaspectratio]{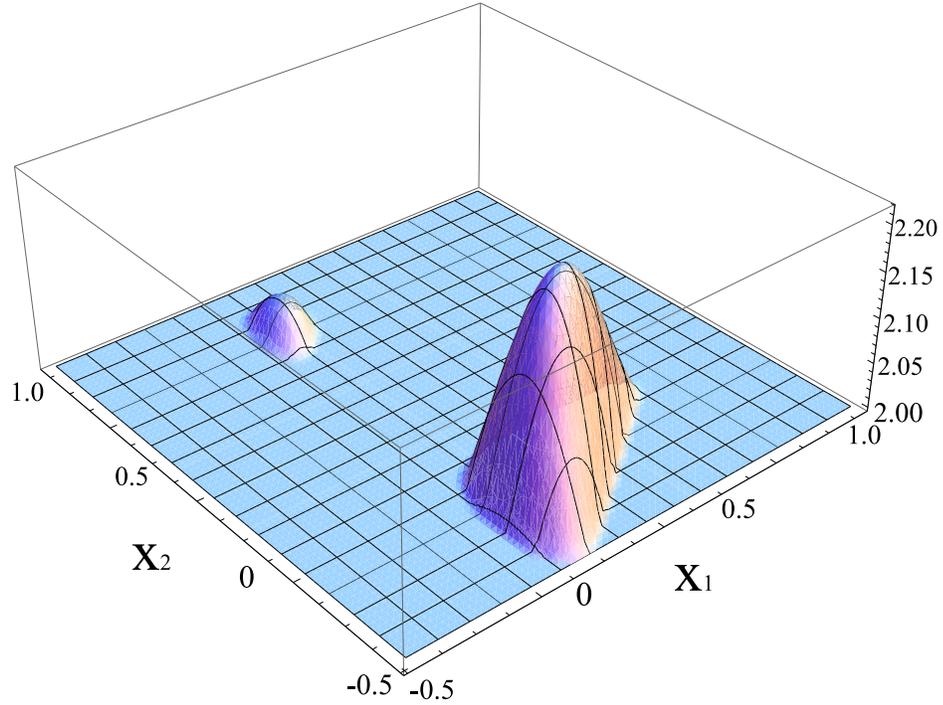}
\end{center}
\caption[fig_5]{The results of calculations of the Bell parameter $B$ in its dependence on the parameters  $x_1,x_2$ for $s_1=0,s_2=9$ and $p_1=p_2=p^{'}_1=p^{'}_2=0, x^{'}_1=0.0164,x^{'}_2=-0.0827$}
\label{fig5}
\end{figure}

\begin{center}
{\bf 5. Conclusion}
\end{center}

Thus, it is demonstrated in the work that, in a strong two-mode electromagnetic field interacting with an atom, the large quantum entanglement may occur. Concerning the value of such entanglement, it can reach such values as  $K\sim 10^8$ and larger. The choice of such parameters is quite simple: it is necessary that the parameter $\eta$ is close to one. For this to happen, it is necessary that the volumes of the resonators $V_1\approx V_2\approx V$ are close to each other and  $\omega \Delta \omega V $ is not large (the smaller the better). The experimental realization of such entangled states is connected, for example, with the crossing of two strong single-mode laser fields, for which the above described conditions are fulfilled.

It should also be said that in the work we have mainly considered the Schmidt parameter for elastic scattering and qualitatively shown that inelastic processes cannot greatly reduce the quantum entanglement. The obtained exact analytical solution of the Schr{\"o}dinger equation for these fields can be used not only for the analysis of entangled states, but also in general to analyze the scattering and inelastic processes in atoms and molecules. Obviously, in such fields, where $\eta$ is close to one, the known approaches to the calculation of inelastic processes such as photoionization and excitation of electrons in atoms and molecules will not work. Indeed, for example, from the expression (\ref{32}) it is seen that the scattered photons can affect the inelastic processes. Usually, in the semi-classical theories, the external classical electromagnetic field is given and determines all the processes in the system, whereas the scattered field does not affect the inelastic processes.

In conclusion, it must be said that the effect of almost complete quantum entanglement of the system for the resonator volumes  $V_1\approx V_2\approx V$ and not large values of the dimensionless parameter $\omega \Delta \omega V $ in the interaction of the two-mode strong electromagnetic field with an atom or molecule is novel and may give impetus to the creation of superpower sources of quantum-entangled photons.

\def\theequation{A\arabic{equation}}
\setcounter{equation}{0}

{\bf Appendix A}

Consider the stationary Schr{\"o}dinger equation with the Hamiltonian (\ref{5}). Let us represent the considered differential equation in the form
\begin{eqnarray}
{\hat{H}}^{'} \Psi^{'}=E\Psi^{'} ,
\label{7}
\end{eqnarray}
where ${\hat{H}}^{'}={\hat{S}}{\hat{H}}{\hat{S}}^{-1}$, $\Psi^{'} = {\hat{S}}\Psi$, while the operator  ${\hat{S}}$ is a certain unitary operator, ${\hat{S}}^{-1} $ is the operator inverse to ${\hat{S}}$ ( ${\hat{S}}^{-1}{\hat{S}}=1 $). Having obtained the solution of the Schr{\"o}dinger equation (\ref{7}), we arrive at the desired wave function by means of the inverse transform $\Psi = {\hat{S}}^{-1}\Psi^{'}$. Next, we introduce a system of coordinates, where we direct a certain vector $\bf S$ along the axis $z$ where in the plane  $xOz$ there will lie a vector $\bf \Sigma$, where $\bf S $ and $\bf \Sigma$ are vectors to be defined below. The solution will be sought for in the form in which the operator
\begin{eqnarray}
{\hat{S}}=exp\left\lbrace i\xi_{2}\frac{\partial }{\partial {x}}\frac{\partial }{\partial {q_{2}}}+i\lambda_{1}\frac{\partial }{\partial {z}}\frac{\partial }{\partial {q_{2}}} + i\xi_{1} \frac{\partial }{\partial {z}}\frac{\partial }{\partial {q_{1}}}\right\rbrace exp\left\lbrace -\lambda_{2}x\frac{\partial }{\partial {z}} \right\rbrace \times
\nonumber\\ 
\times exp\left\lbrace \xi q_{2}\frac{\partial }{\partial {q_{1}}} \right\rbrace exp\left\lbrace -\eta q_{1}\frac{\partial }{\partial {q_{2}}} \right\rbrace .
\label{8}
\end{eqnarray}
It should be noted that the choice of the operator ${\hat{S}}$ in the form  (\ref{8}) is made as a result of careful analysis of the stationary Schr{\"o}dinger equation with the Hamiltonian (\ref{5}) in terms of the possibility of its diagonalization. In (\ref{8}) the constant quantities  $\xi_{1}, \xi_{2},\xi,\eta, \lambda_{1},\lambda_{2} $ are unknown, and we find them in further consideration. Now our goal is to find  $\xi_{1}, \xi_{2},\xi,\eta, \lambda_{1},\lambda_{2} $ such that the Hamiltonian  ${\hat{H}}^{'}$ in (\ref{7}) becomes diagonal. To do this, we introduce an auxiliary Hamiltonian
\begin{eqnarray}
{\hat{\tilde{H}}}= exp\left\lbrace \xi q_{2}\frac{\partial }{\partial {q_{1}}} \right\rbrace exp\left\lbrace -\eta q_{1}\frac{\partial }{\partial {q_{2}}} \right\rbrace {\hat{H}} exp\left\lbrace \eta q_{1}\frac{\partial }{\partial {q_{2}}} \right\rbrace exp\left\lbrace -\xi q_{2}\frac{\partial }{\partial {q_{1}}}\right\rbrace .
\label{9}
\end{eqnarray}
Let us find such values of  $ \xi,\eta $ in $ \hat{\tilde{H}} $, that there will be no "crossing" values for  $q_{1}$ and $q_{2}$ (the product  $q_{1} q_{2}$ disappears). Carrying out the necessary calculations, we get
\begin{eqnarray}
{\hat{\tilde{H}}}= -\frac{1}{2} \Delta -i q_{1}{\bf{\nabla S}}-i q_{2}{\bf{\nabla \Sigma}}+q^2_{1} W^2+q^2_{2}\Lambda^2 - V^2\frac{\partial^2 }{\partial {q^2_{1}}} - L^2\frac{\partial^2 }{\partial {q^2_{2}}},
\label{10}
\end{eqnarray}
where
\begin{eqnarray}
{\bf{\Sigma}}=\frac{1}{2}\sqrt{\frac{\omega_{1}}{\omega_{2}}}\frac{\gamma\beta_{1}}{\sqrt{\alpha^2+\gamma^2}}{\bf{u}}_{1}+\frac{\beta_{2}}{2}\left( 1+\frac{\alpha}{\sqrt{\alpha^2+\gamma^2}}\right){\bf{u}}_{2} ~, 
\nonumber\\ 
{\bf{S}} =\beta_{1} {\bf{u}}_{1}-\sqrt{\frac{\omega_{2}}{\omega_{1}}}\beta_{2}\frac{\sqrt{\alpha^2+\gamma^2}-\alpha}{\gamma}{\bf{u}}_{2} ~,
\nonumber\\ 
W^2=\frac{\omega_{2}}{\omega_{1}}(\omega_{2}+\beta^2_{2})\left(1+\left( \frac{\alpha}{\gamma}\right)^2 - \frac{\alpha \sqrt{\alpha^2+\gamma^2}}{\gamma^2} \right) - \frac{ \sqrt{\alpha^2+\gamma^2}}{2},
\label{11}
\nonumber\\ 
\Lambda^2 = \frac{1}{4}(\omega_{2}+\beta^2_{2})\left(1+\frac{\alpha}{\sqrt{\alpha^2+\gamma^2}} \right)+\frac{1}{8}\frac{\omega_{1}}{\omega_{2}}\frac{\gamma^2}{\sqrt{\alpha^2+\gamma^2}} ~,
\nonumber\\
V^2 = \frac{\omega_{1}}{4}\left(1+\frac{\alpha}{\sqrt{\alpha^2+\gamma^2}} \right),~~~~ L^2 = \omega_{2}\left(1 + \left( \frac{\alpha}{\gamma}\right)^2 \left(1-\frac{ \sqrt{\alpha^2+\gamma^2}}{\alpha} \right)  \right).  
\end{eqnarray}
In the expression (\ref{11})
\begin{eqnarray}
\alpha = \frac{\omega_{2}}{\omega_{1}}(\omega_{2}+\beta^2_{2})-(\omega_{1}+\beta^2_{1}),~~~~\gamma = 2\sqrt{ \frac{\omega_{2}}{\omega_{1}}}\beta_{1}\beta_{2} {\bf{u}}_{1} {\bf{u}}_{2}.
\label{12}
\end{eqnarray}
Besides, in finding (\ref{10}) the parameters 
\begin{eqnarray} 
\xi = \frac{1}{2}\sqrt{\frac{\omega_{1}}{\omega_{2}}}\frac{\gamma}{\sqrt{\alpha^2+\gamma^2}},~~~~ \eta = \sqrt{\frac{\omega_{2}}{\omega_{1}}}\frac{\sqrt{\alpha^2+\gamma^2}-\alpha}{\gamma} .
\label{13}
\end{eqnarray}
Next, we find the Hamiltonian 
\begin{eqnarray} 
{\hat{H}}^{'}=exp\left\lbrace i\xi_{2}\frac{\partial }{\partial {x}}\frac{\partial }{\partial {q_{2}}}+i\lambda_{1}\frac{\partial }{\partial {z}}\frac{\partial }{\partial {q_{2}}} + i\xi_{1} \frac{\partial }{\partial {z}}\frac{\partial }{\partial {q_{1}}}\right\rbrace exp\left\lbrace -\lambda_{2}x\frac{\partial }{\partial {z}} \right\rbrace 
{\hat{\tilde{H}}}\times
\nonumber\\ 
\times exp\left\lbrace i\lambda_{2}x\frac{\partial }{\partial {z}} \right\rbrace exp\left\lbrace -i\xi_{2}\frac{\partial }{\partial {x}}\frac{\partial }{\partial {q_{2}}}-i\lambda_{1}\frac{\partial }{\partial {z}}\frac{\partial }{\partial {q_{2}}} - i\xi_{1} \frac{\partial }{\partial {z}}\frac{\partial }{\partial {q_{1}}}\right\rbrace . 
\label{14}
\end{eqnarray}
Choosing the unknown parameters  $\xi_{1}, \xi_{2}, \lambda_{1},\lambda_{2} $ in such a way that the Hamiltonian ${\hat{H}}^{'}$ becomes diagonal, we get 
\begin{eqnarray}
{\hat{H}}^{'}= -\frac{1}{2} \Delta +q^2_{1} W^2+q^2_{2}\Lambda^2 - V^2\frac{\partial^2 }{\partial {q^2_{1}}} - L^2\frac{\partial^2 }{\partial {q^2_{2}}}+
\nonumber\\ 
+\left( \frac{S^2}{4W^2}+\frac{1}{2}\frac{\left( \frac{{\bf \Sigma S}}{S}\right)^2 }{2\Lambda^2 -\left( \frac{{\bf \Sigma \times S}}{S}\right)^2}\right) \frac{\partial^2 }{\partial {z^2}}+\left( \frac{{\bf \Sigma \times S}}{S}\right)^2\frac{1}{4\Lambda^2}\frac{\partial^2 }{\partial {x^2}} .
\label{15}
\end{eqnarray}
In obtaining the expression (\ref{15})  $\xi_{1}, \xi_{2}, \lambda_{1},\lambda_{2} $ are chosen in the form 
\begin{eqnarray}
\xi_{1} = \frac{S}{2W^2},~~~ \xi_{2} =  \frac{|{\bf \Sigma \times S}|}{2S \Lambda^2},~~~\lambda_{1}=\frac{{\bf \Sigma  S}}{S}\frac{1}{2\Lambda^2 -\left( \frac{{\bf \Sigma \times S}}{S}\right)^2},~~~ \lambda_{2}=\frac{1}{S^2}\frac{|{\bf \Sigma \times S}|{\bf \Sigma  S}}{2\Lambda^2 -\left( \frac{{\bf \Sigma \times S}}{S}\right)^2} .
\label{16}
\end{eqnarray}
It is not difficult to solve the Schr{\"o}dinger equation with the Hamiltonian (\ref{15}), since all variables are separated, and these solutions will be in the form of a plane wave and the wave functions of the harmonic oscillator. First we write an eigenvalue of the energy of the Hamiltonian (\ref{15})
\begin{eqnarray}
E_{{\bf k},m,n}=\frac{k^2_{x}}{2}\left(1- \frac{\left( {\bf \Sigma \times S}\right)^2 }{2S^2 \Lambda^2}\right) + \frac{k^2_{z}}{2}\left(1-\left(\frac{S^2}{2W^2} + \frac{1}{S^2}\frac{\left( {\bf \Sigma  S}\right)^2 }{2\Lambda^2 -\left( \frac{{\bf \Sigma \times S}}{S}\right)^2}\right) \right) +
\nonumber\\ 
+ \frac{k^2_{y}}{2}+2VW\left( m+\frac{1}{2}\right) +2L\Lambda\left( n+\frac{1}{2}\right),  ~~~~~
\label{17}
\end{eqnarray}
where $k_{x},k_{y},k_{z}$ are the projections of the wave vector  $\bf k$ of the free particle on the corresponding coordinate axes, whereas $n,m = 0,1,2,...$ are quantum numbers. Next, we write the Eigen wave function of the Hamiltonian (\ref{15})
\begin{eqnarray}
\Psi^{'}_{{\bf k},m,n}= C_{\bf k}e^{i{\bf k}{\bf r}}C_{n}e^{-\frac{\Lambda}{L}\frac{q^2_{2}}{2}}H_{n}\left(q_{2}\sqrt{\frac{\Lambda}{L}} \right) C_{m}e^{-\frac{W}{V}\frac{q^2_{1}}{2}}H_{m}\left(q_{1}\sqrt{\frac{W}{V}} \right),
\label{18}
\end{eqnarray}
where $ C_{\bf k}$ is the normalization factor for the wave function of an electron,  $H_{n}$ are the Hermite polynomials, whereas the normalizing wave functions for the electromagnetic fields
\begin{eqnarray}
C_{n}=\frac{1}{\sqrt{2^n n!\sqrt{\pi}}} \left( \frac{\Lambda}{L}\right)^{-1/4} ,~~~~~~ C_{m}=\frac{1}{\sqrt{2^m m!\sqrt{\pi}}} \left( \frac{W}{V}\right)^{-1/4}.
\label{19}
\end{eqnarray}
To find the wave function $\Psi$, we need  ${\hat{S}}^{-1}$, which is known, since all the parameters in the operator  ${\hat{S}}$ are known. By acting by the operator ${\hat{S}}^{-1}$ on the wave function $\Psi^{'}$, it is quite simple to obtain the desired wave function. Indeed, after the action of ${\hat{S}}^{-1}$ on the wave function of the electron, we obtain the shift operators. As a result, we get (\ref{20}).

\def\theequation{B\arabic{equation}}
\setcounter{equation}{0}

{\bf Appendix B}

Consider integral (\ref{22}),  where $\Psi_{{\bf k},m,n}$ is determined by the expression (\ref{24}). It should be said that this integral cannot be reduced to the standard ones; therefore, we consider it in detail. To calculate it, we represent $\Psi_{{\bf k},m,n}$ in the form 
\begin{eqnarray}
\Psi_{{\bf k},m,n}= C_{\bf k}e^{i{\bf k}{\bf r}}\sum_{k,p}G^{In}_{m,n,k,p}\Psi^{El}_{k,p,s_1,s_2} ,
\label{44}
\end{eqnarray}
where $\Psi^{El}_{k,p,s_1,s_2}$ is the wave function of two-mode electromagnetic field in the elastic scattering
\begin{eqnarray}
\Psi^{El}_{k,p,s_1,s_2}= C_{k}C_{p}exp\left( -\frac{\sigma}{2}\left(q_{2}+\eta q_{1}\right) ^2\right) H_{p}\left(\sqrt{\sigma}\left(q_{2}+\eta q_{1} \right) \right)
\nonumber\\ 
\times
exp\left( -\frac{\sigma}{2}\left( q_{1}-\eta q_{2}\right)^2\right) H_{k}\left(\sqrt{\sigma}\left( q_{1}-\eta q_{2} \right) \right) ,
\label{45}
\end{eqnarray}
where $C_{k}, C_{p}$ are normalization constant (\ref{19}). Next, we find the decomposition coefficient $G^{In}_{m,n,k,p}$. To this end, we multiply expression (\ref{44}) by the wave function  (\ref{45}) and integrate using the orthogonality condition $\langle \Psi^{El}_{k,p,s_1,s_2}|\Psi^{El}_{k^{'},p^{'},s_1,s_2}\rangle =\delta_{k,k^{'}}\delta_{p,p^{'}}$ ($\delta_{a,b}$ is the Kronecker symbol). The resulting integral can be calculated rather easily by reducing it to a standard one  \cite{Prudnikov_T3}
\begin{eqnarray}
\int^{\infty}_{-\infty}e^{-Q^2} H_{n}(Q+\Omega)H_{p}(Q-\Omega)dQ=2^p\sqrt{\pi}n!(-\Omega)^{p-n}L^{p-n}_{n}(2\Omega^2),
\label{46}
\end{eqnarray}
where $L^{b}_{a}(x)$ is the generalized Laguerre polynomial, while this expression for the integral holds for $p\geq n$. If $p < n$, then in the expression we must substitute $n$ by $p$, $p$ by $n$, and also replace $\Omega$ by $-\Omega$. As a result, we obtain the expression (\ref{27}). Next, we calculate the integral (\ref{22}) substituting in it the expression (\ref{44}); finally, we get $A^{s_1,s_2,0}_{m,n,{\bf k}}=F_{{\bf k},0}\sum^{\infty}_{k,p}G^{In}_{m,n,k,p}({\bf k})G^{El}_{k,p,s_{1},s_{2}}$, where
\begin{eqnarray}
G^{El}_{k,p,s_{1},s_{2}}= C_{k}C_{p} C_{s_1}C_{s_2}\int e^{\left( -\frac{\sigma}{2}\left(q_{2}+\eta q_{1}\right) ^2\right)} H_{p}\left(\sqrt{\sigma}\left(q_{2}+\eta q_{1} \right) \right)
\nonumber\\ 
\times
e^{\left( -\frac{\sigma}{2}\left( q_{1}-\eta q_{2}\right)^2\right)} H_{k}\left(\sqrt{\sigma}\left( q_{1}-\eta q_{2} \right) \right)e^{-\frac{1}{2}q^2_1}e^{-\frac{1}{2}q^2_2}H_{s_1}(q_1)H_{s_2}(q_2) dq_1 dq_2.
\label{47}
\end{eqnarray}
The expression (\ref{47}) is not a standard integral. We calculate it using the definition of the Hermite polynomials. We conclude that this integral can be represented in the form 
\begin{eqnarray}
G^{El}_{k,p,s_{1},s_{2}}= C_{k}C_{p} C_{s_1}C_{s_2}(-1)^{s_1+s_2+p+k}\frac{d^{p}}{dx^p} \frac{d^{k}}{dy^k}e^{-x^2}e^{-y^2}
\nonumber\\ 
\times 
\int^{\infty}_{-\infty}e^{-2\sqrt{\sigma}q_1(y+\eta x)}\frac{d^{s_1}}{dq^{s_1}_1} e^{-q^2_{1}}dq_1 \int^{\infty}_{-\infty}e^{-2\sqrt{\sigma}q_2(x-\eta y)}\frac{d^{s_2}}{dq^{s_2}_2} e^{-q^2_{2}}dq_2.
\label{48}
\end{eqnarray}
In the expression (\ref{48}) after taking the derivatives with respect to  $x,y$ we should assume  $x=y=0$. It is not difficult to calculate integrals in the expression (\ref{48}); as a result, we get 
\begin{eqnarray}
G^{El}_{k,p,s_{1},s_{2}}= C_{k}C_{p} C_{s_1}C_{s_2}\pi (2\sqrt{\sigma})^{s_1+s_2}\left( \frac{d^{p}}{dx^p} \frac{d^{k}}{dy^k}(y+\eta x)^{s_1}(x-\eta y)^{s_2}\right)\Biggr|_{x=0,y=0} .
\label{49}
\end{eqnarray}
So, we have obtained a polynomial. It can be seen from the expression (\ref{49}) that the condition  $s_1+s_2=k+p$, holds, which means that the number of particles in the elastic scattering is preserved. Using the properties of the Jacobi polynomials, it is not difficult to show that  $G^{El}_{k,p,s_{1},s_{2}}$ will have the form  (\ref{26}).

\end{document}